\documentclass[reprint, amsmath,amssymb, aps, prl, longbibliography, floatfix]{revtex4-1}

\usepackage{graphicx}
\usepackage{dcolumn}

\usepackage{bm}
\usepackage{hyperref}
\usepackage{braket}
\usepackage{longtable}
\usepackage[usenames, dvipsnames]{color}
\usepackage[mathlines]{lineno}

\usepackage{fancyhdr}
\definecolor{mygray}{gray}{0.6}

\begin{document}


\title{Superconducting Through-Silicon Vias for Quantum Integrated Circuits }

\author{Mehrnoosh Vahidpour, William O'Brien, Jon Tyler Whyland, Joel Angeles, Jayss Marshall, Diego Scarabelli, Genya Crossman, Kamal Yadav, Yuvraj Mohan, Catvu Bui, Vijay Rawat, Russ Renzas, Nagesh Vodrahalli, Andrew Bestwick, Chad Rigetti}
\affiliation{Rigetti Computing, 775  Heinz  Avenue,  Berkeley,  CA  94710}


\date{\today}

\begin{abstract}
We describe a microfabrication process for superconducting through-silicon vias appropriate for use in superconducting qubit quantum processors. With a sloped-wall via geometry, we can use non-conformal metal deposition methods such as electron-beam evaporation and sputtering, which reliably deposit high quality superconducting films. Via superconductivity is validated by demonstrating zero via-to-via resistance below the critical temperature of aluminum.
\end{abstract}

\maketitle

\section{Introduction}

Through-silicon vias (TSVs) have been widely used in semiconducting integrated devices in recent years, both as high-performance interconnects to create 3D circuits and to engineer isolation between components on chips \cite{overview,TSVbook}. Superconducting qubit circuits, composed of microwave resonant structures that store and manipulate quantum information, will require similar technical solutions as the number of qubits grow. A truly scalable quantum integrated circuit architecture requires both 3D integration and careful engineering of the RF environment in which qubits operate.

\begin{figure}
\includegraphics[width=0.45\textwidth]{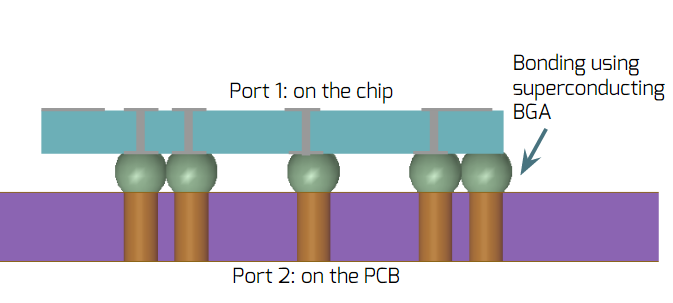}
\caption{An example architecture of a 3D integrated quantum circuit. A chip with a tiled qubit design can be bonded to a printed circuit board on the backside using superconducting ball grid array (BGA). The insertion loss could be quite low for such a setup.
\label{signaldeliverysetup}}
\end{figure}

The advantages of using TSVs in quantum integrated circuits are:

\textit{Signal delivery.} Traditionally, signal delivery in superconducting qubit devices has been in the plane of the circuit through bond pads at the boundary of the chip. However, most proposals for large quantum circuits involve tiled 2D lattices in which interior qubits cannot be accessed from the perimeter {\cite{surfacecode}}. Delivering signals perpendicular to the plane of the 2D circuit is therefore necessary for future scalability. Flip-chip processes using superconducting indium bumps have recently been used to bond a chip with qubits to another chip with readout and control signals demonstrating high qubit coherence ($>$20 $\mu$s) \cite{lincoln, google,google2}. 3D integration using vias, on the other hand, enables delivering signals through the substrate. A simple schematic of  such an architecture is shown in Fig.~\ref{signaldeliverysetup} in which high frequency signals travel from an interposer to the back side of the chip, through the vias, onto the top surface, and to the qubits. 

\textit{Isolation.} To achieve maximum coherence and minimal crosstalk, the resonant elements of a quantum integrated circuit should ideally be completely isolated from each other and from the environment, except for engineered couplings. This would also eliminate frequency-crowding concerns for large circuits. TSVs can partially accomplish this if arrange so as to surround resonant elements, locally confining electromagnetic modes in the substrate. A conductive cage above the plane of the chip would complete the creation of a 3D enclosure.

\textit{Suppressing substrate modes.} Dielectric substrates host electromagnetic modes. A thin chip with dimensions $x \times y$ has a fundamental mode with frequency $(c/2\pi\sqrt{\epsilon_r})\sqrt{(\pi/x)^2+(\pi/y)^2}$ , where $c$ is the speed of light in free space and $\epsilon_r$  is the dielectric constant. For sufficiently large substrates, the  mode falls in the operating range of a superconducting qubit device, typically 3-10 GHz, and provides a loss channel for otherwise high-quality resonators. In practice, this places an upper bound on circuit size and complexity one can build on a single chip. However, the presence of a large number of metal TSVs in the substrate enforces boundary conditions on substrate modes, effectively limiting the maximum wavelength to the TSV spacing. This can raise the lowest-lying mode frequency well above 10 GHz and eliminate its dependence on substrate size, allowing for arbitrarily large chips.

\textit{Enabling coupling between layers.} Long-term scaling of quantum integrated circuits will require multi-layer structures. In such a geometry, the precision of the vertical distance between layers is usually set by the fabrication technology, making capacitative coupling between layers difficult to engineer. Rather, a galvanic connection using vias is preferable for delivering signals between different planes, and coupling capacitances could be designed within planes.

\section{TSV Manufacturing in the Semiconductor Industry}

The state of the art of TSV manufacturing in the semiconductor industry typically includes the following process steps \cite{TSVbook}:

\textit{Via formation:} Deep reactive ion etching or laser drilling to define the via shape.

\textit{Dielectric barrier deposition:} Thermal silicon oxide or nitride, or plasma-enhanced chemical vapor deposition (PECVD) of a thin film to form  diffusion barrier between Cu and Si.

\textit{Adhesion and seed layer deposition:} Physical vapor deposition (PVD) of Ti, followed by Cu.

\textit{Cu electroplating:} Fill the via with conductive Cu. Alternative electroplating metals such as W and Sn are also used. Usually, it is followed by an annealing process to release stress.

\textit{Chemical mechanical polishing (CMP):} Removal of Cu overburden.

\textit{TSV reveal:} Mechanical and chemical grinding or polishing (hard reveal) or wet etching (soft reveal) of the wafer to reveal the vias.

This process, while manufacturable and widely used for CMOS circuits, could be incompatible with superconducting qubits for three reasons. First, the dielectric liner would be a site of microwave loss. Second, the normal metal fill will introduce ohmic heating and loss, whether due to signals or grounding currents. Third, poorly controlled CMP processes could lead to small levels of contamination and doping in otherwise insulating Si substrates. All three factors would reduce qubit lifetimes.

Furthermore, these problems cannot be resolved with simple materials substitutions. There is no well-established process for electroplating of superconducting metals such as Al. High-quality superconducting microwave structures are typically fabricated via physical deposition, but such methods deposit metal too slowly and with too much stress to fill deep holes. Even worse, the difference in thermal expansion properties between Si and a metal fill may prohibit cryogenic use of these structures.

\section{TSV manufacturing in quantum integrated circuits}

We propose a process for manufacturing superconducting TSVs which uses a superconducting metal as the via liner, a polymer for partial filling, and avoids the need for CMP by employing a temporary membrane to support the via liner.

\textit{Support membrane.} Unlike industry processes in which thick wafers are etched and later polished to reveal the vias, we etch entirely through the Si wafer during via formation. First, however, we deposit a film on the back of the wafer which is highly selective to our silicon etch and therefore acts as stop. In subsequent steps it supports the deposition of liners and fills. At the end of the process it is easily etched away, revealing a pristine Si surface with exposed metal tops of vias, onto which the superconducting circuit can be fabricated.

\textit{Etch geometry and deposition method.} In our process, via walls are etched to be slanted, which allows us to substitute electroplating for more directional deposition methods such as electron-beam evaporation, epitaxy, and sputtering. For a truly directional process, we can easily calculate the coverage on the sidewalls as a function of both via sidewall angle and deposition angle.

\begin{figure}
\includegraphics[width=0.5\textwidth]{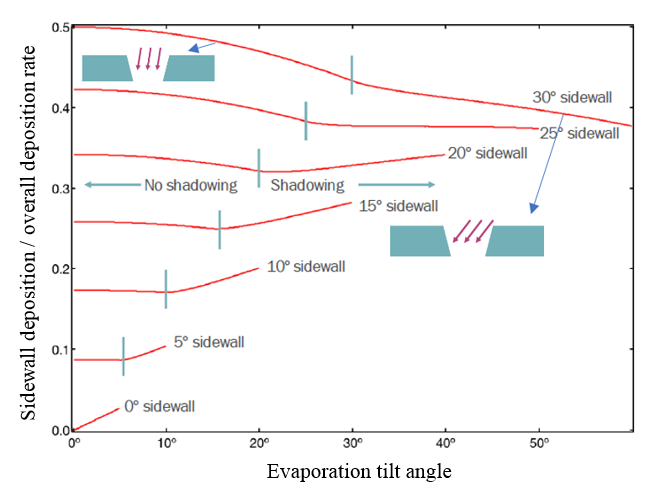}
\caption{Simulated sidewall deposition rate (normalized to the deposition rate onto a perpendicular surface) as a function of sidewall angle and deposition angle.
\label{EvapAngle}}
\end{figure}

Fig.~\ref{EvapAngle} shows the ratio of the sidewall deposition rate to the overall deposition rate on a normal surface as a function of the evaporation angle for different sidewall tapers. For evaporation angles smaller than the TSV angle, both sidewalls are coated, while at least a portion of the sidewalls is shadowed when the evaporation angle is larger than the TSV angle. In the former regime, a nonzero deposition angle reduces sidewall deposition. Hence, we primarily deposit onto our sloped walls at an angle normal to the plane of the wafer, with no substrate tilt relative to the deposition source.

The choice of the via sidewall angle depends on: 1) the minimum required thickness of the metal on the walls; 2) the maximum possible deposition thickness; and 3) the maximum allowed difference between the via diameter at each end (which depends on the minimum required via density and the maximum acceptable via volume).

Based on the above parameters, sidewall angles in the range of 10 to 20 degrees are small enough to allow acceptable difference between the opening on the sides of the wafers (in our case, $\sim 150\mu$m) and at the same time large enough to allow sidewall coverage in the range of 20-30$\%$ of the amount deposited on a normal surface.

\textit{Polymer fill.} It is still necessary to fill the via to provide permanent support for the thin metal liner once the membrane is removed. The criteria for the selection of fill materials are: 1) a straightforward deposition process that does not alter the thin membrane and liner film; 2) mechanical strength to provide permanent support to the via metal liner; 3) robustness to downstream chemical processing; 4) cyrogenic compatibility, including thermal expansion properties similar to silicon. The microwave properties of the material are not relevant as long as the superconducting liner is thick enough to completely shield it from the circuit.

Thick polymers are one promising class of materials for this purpose, among which Parylene-C was chosen \cite{Parylene_C}. It has the desired mechanical and chemical properties, is inert to almost all wet process, and has a simple deposition process. Due to its conformal, pinhole-free, and easily controlled deposition process (even for ultra-thin coatings), it is commonly used to protect electronic components and compact packages as well as sensors, LEDs, and MEMS \cite{Parylene_C2}. We have also tested the mechanical and cryogenic properties with no sign of degradation after repeated thermal cycles.

\section{Superconducting TSV process}
Here, we describe the process we have developed for fabricating grounded superconducting vias for isolation and for suppressing substrate modes. In this process all the vias are grounded through a blanket metal deposition on the back side side of the wafer. Future use for 3D signal delivery will require additional process steps to pattern the wafer backside. The process steps are as follows (Fig.~\ref{fabprocess}):

\begin{figure}
\includegraphics[width=0.45\textwidth]{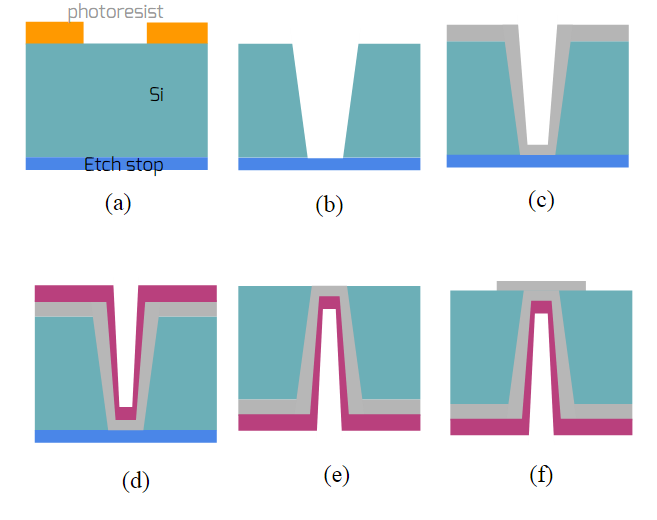}
\caption{Schematic of our fabrication process for superconducting Al vias. (a) Silicon oxynitride membrane deposition (for etch stop) and thick photoresist patterning of the vias, (b) Via etch, (c) Al electron-beam evaporation. (d) Parylene-C physical vapor deposition,  (e) Membrane removal / via reveal, (f) Device side processes.
\label{fabprocess}}
\end{figure}

\textit{Starting substrate.}
For superconducting microwave structures, we use 300 $\mu$m-thick high-resistivity ($>$15 k$\Omega\cdot$cm) silicon wafers. First, we deposit an etch stop for the via etch process. We perform PECVD of a silicon oxynitride SiO$_x$N$_y$ film with the thickness of 7-10 $\mu$m. The stoichiometry is tuned to minimize stress and wafer bow, reduce stress changes in the membrane upon etch stop removal, and be easily removed using wet etch chemistries. Since the etch is deep, a thick positive photoresist ($>$ 10 $\mu$m) is used to pattern the vias on the non-oxynitride side of the wafer.

\begin{figure}
\includegraphics[width=0.35\textwidth]{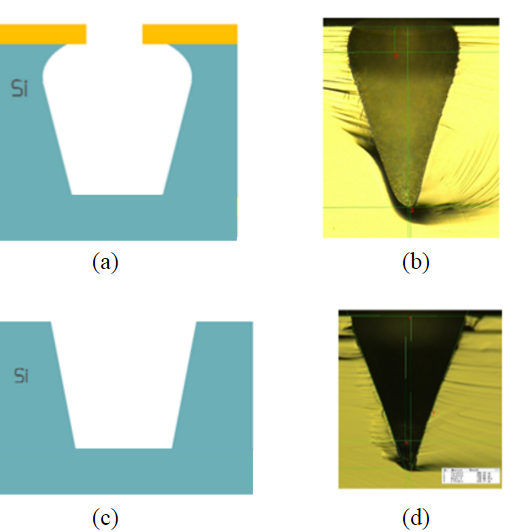}
\caption{(a) Via main etch process, and (b) actual etch on a bulk test silicon wafer. An overhang is created during the main etch process. (c) Via global isotropic etch after stripping the photoresist, and (d) actual etch on a bulk test silicon wafer. The overhang is removed in this step.
\label{ViaEtch}}
\end{figure}

\textit{Via formation.}
The via formation process employs the method described in \cite{SF6_O2,SF6_O2_4} where a non-Bosch process in an SPTS inductively coupled plasma (ICP) tool is used to etch the desired via taper. The continuous plasma etching process uses a mixture of SF$_6$/O$_2$/Ar gases to etch vias with angles in the range of  10 to 20 degrees. The etch process is performed in two steps, which we refer to as the main via etch and the isotropic via etch. In the main etch, the sloped vias are formed with an overhang as a result of the etch undercut (Fig.~\ref{ViaEtch}a-b). Next, in the isotropic etch step, we strip off the photoresist and perform a global etch to remove the overhang (Fig.~\ref{ViaEtch}c-d).

Approximately 50 $\mu$m of the wafer thickness is etched away during the isotropic etch step, making the final wafer thickness $\sim$250 $\mu$m. Our target angle is 10-20 degrees, and is achieved with SF$_6$/O$_2$ flows of 135/70 sccm and 300/75 sccm for the main and iso etch steps, respectively. Also, since the etch stops at an isolating membrane, to avoid footing (which is a common problem for dry silicon etching in ICP processes) we employ a low frequency pulsed power for the platen.

\textit{Superconducting liner deposition.}
In this step we coat the via walls with Al using electron-beam evaporation. For the film to effectively shield radiation it should be several times larger than the London penetration depth ($\sim$50 nm). A 2.5 $\mu$m deposition guarantees minimum of 250 nm deposition on the sidewalls for angles above 10 degrees, based on the calculations represented in Fig.~\ref{EvapAngle}. This blanket deposition coats the walls of the vias as well as the entire backside of the wafer, and grounding all vias together.

\begin{figure}
\includegraphics[width=0.35\textwidth]{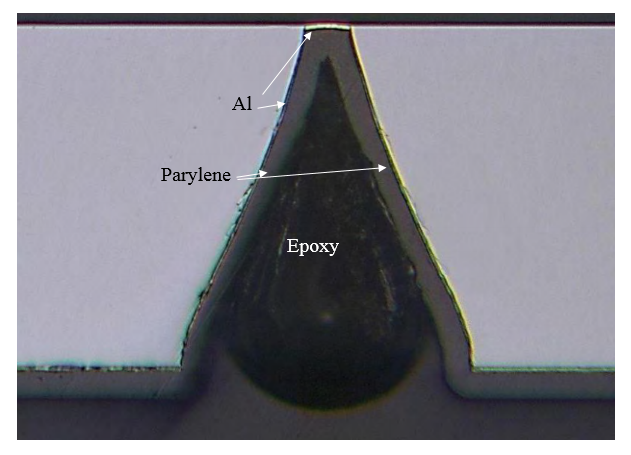}
\caption{Cross section of the via after Parylene-C deposition. Note that the epoxy resin is used to fill the vias for cross-sectioning and imaging purposes.
\label{parylene}}
\end{figure}

\textit{Via fill.}
We deposit a 20 $\mu$m layer of Parylene-C film in a PVD process. It partially fills the vias on the sidewalls and the bottom, while leaving most of the via volume empty.  Fig.~\ref{parylene} shows the cross-section image of the vias after Parylene deposition step.

\begin{figure}
\includegraphics[width=0.25\textwidth]{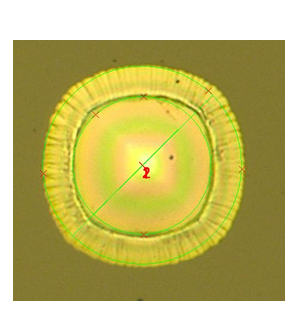}
\caption{Top view of a 50 um diameter via after the reveal. Two concentric circles are shown which are the representative of the main and isotropic etch processes.
\label{AfterReveal}}
\end{figure}

\textit{Via reveal.} Finally, the oxynitride film is etched using buffered HF acid. The via Al is revealed after this process (see Fig.~\ref{AfterReveal} for a top view). The two concentric circles shown on the revealed side of the via are representative of the main and isotropic etch process, which cause different surface roughnesses on the silicon and the oxynitride membrane. This roughness is transferred to the deposited Al film.

\begin{figure}
\includegraphics[width=0.45\textwidth]{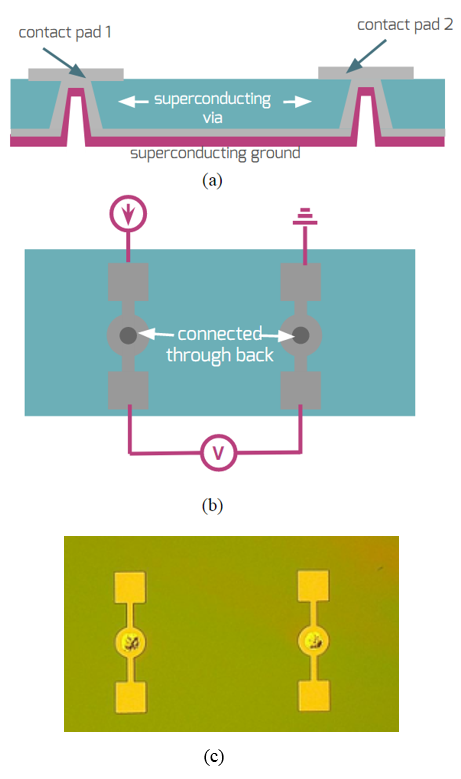}
\caption{4-point resistance measurement setup (a) side view (b) top view for a via pair. (c) Photograph of the fabricated device for this test.
\label{4pointsetup}}
\end{figure}
After the reveal step, the wafer is ready for the quantum circuit fabrication processes.

\section{ Superconductivity Measurement }

To verify the superconductivity of the vias, we measured DC resistances between pairs of vias in a dilution refrigerator. Since all vias on a given wafer are connected through the back side, we can measure in a 4-terminal arrangement the voltage drop resulting from a current flowing from the front side, through one via, across the back plane, and returning to the front side through a second via (Fig.~\ref{4pointsetup}a-b).

\begin{figure}
\includegraphics[width=0.45\textwidth]{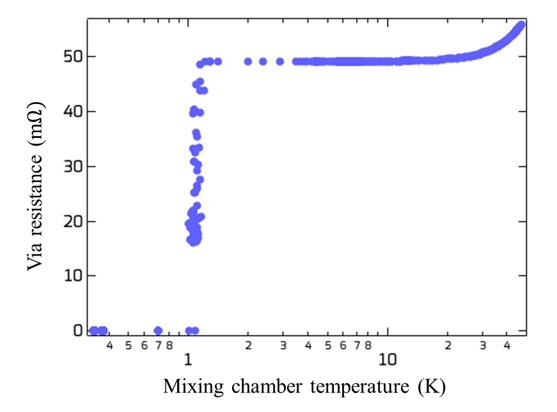}
\caption{4-point via-to-via resistance measurement as a function of temperature. The resistance of the via pair drops to zero at temperatures below 1.2 K, near the expected Al critical temperature.
\label{4pointmeasurement}}
\end{figure}

To prepare the sample, we patterned contact pads, removed the native oxide on the surface of the via Al with a 400 V, 20 mA argon ion mill etch, and deposited \textit{in situ} a 160 nm Al film using electron-beam evaporation. After liftoff, the test structures appear as in Fig.~\ref{4pointsetup}c.

We measured the via-to-via resistance with a lock-in amplifier while cooling the sample to subkelvin temperatures (Fig.~\ref{4pointmeasurement}). As expected, the resistance drops to zero below the critical temperature of Al, around 1.2 K. Our instrumentation sets the lower bound of the critical current of this device to be 100 $\mu$A.

\section{Conclusions}
A microfabrication process for superconducting TSV with Al, sloped via walls and Parylene-C as the via fill was developed. The via geometry allows using non-conformal metal deposition methods such as e-beam evaporation. We demonstrated the superconductivity of vias by measuring the zero via pair resistance below the critical temperature of Al.

This establishes the viability of the process for making grounded superconducting vias. Future work is required to demonstrate compatibility with superconducting qubit fabrication and measurement, as well as the extra process steps required to use vias to deliver microwave signals.

\end{document}